\documentclass[letterpaper, 10 pt, conference]{ieeeconf}  

\IEEEoverridecommandlockouts                              

\let\labelindent\relax

\usepackage[dvipsnames]{xcolor}

\newcommand{\vo}[1]{\boldsymbol{#1}}

\usepackage{lipsum}

\usepackage{graphicx,cite} 
\usepackage{amsmath, enumitem,amssymb} 
\usepackage{empheq} 

\title{\LARGE \bf
$\mathcal{H}_2$ Optimized PID Control of Quad-Copter Platform \\ with Wind Disturbance}

\author{Sunsoo Kim$^{1}$ and Vedang Deshpande$^{2}$ and Raktim Bhattacharya$^{3}$
\thanks{$^{1}$Sunsoo Kim is a Ph.D student in the Department of Electrical and Computer Engineering,  Texas A\&M University, College Station, TX 77840, USA. Email: {\tt\small kimsunsoo@tamu.edu}}%
\thanks{$^{2}$ Vedang Deshpande is a Ph.D student in the Department of Aerospace Engineering, Texas A\&M University, College Station, TX 77840, USA. Email:
        {\tt\small vedang.deshpande@tamu.edu}}%
\thanks{$^{3}$Raktim Bhattacharya is with the Faculty of Aerospace Engineering, Texas A\&M University, College Station, TX 77840, USA. Email:
        {\tt\small raktim@tamu.edu}}%
}

\begin{document}
\maketitle
\thispagestyle{empty}
\pagestyle{empty}

\begin{abstract}
Proportional-Integral-Derivative (PID) scheme is the most commonly used algorithm for designing the controllers for unmanned aerial vehicles (UAVs). However, tuning PID gains is a non trivial task. A number of methods have been developed for tuning the PID gains for UAV systems. However, these methods do not handle wind disturbances, which is a major concern for small UAVs. In this paper, we propose a new method for determining optimized PID gains in the $\mathcal{H}_2$ optimal control framework, which achieves improved wind disturbance rejection. The proposed method compares the classical PID control law with the $\mathcal{H}_2$ optimal controller to determine the $\mathcal{H}_2$ optimal PID gains, and involves solving a convex optimization problem. The proposed controller is tested in two scenarios, namely, vertical velocity control, and vertical position control. The results are compared with the existing LQR based PID tuning method.

\end{abstract}

\section{INTRODUCTION} \label{sec:intro}
In recent years, unmanned aerial vehicles (UAVs) have found applications in many diverse fields encompassing commercial, civil, and military sectors \cite{mazur2016pwc,canetta2017exploring,Droneii}. Because of their vertical take-off and landing capabilities and relative simplicity in modeling, quadcopters have become one of the most popular choice for UAVs, and a number of algorithms have been developed to control them \cite{zulu2016review}.

Among these algorithms, PID control is still the most popular algorithm in the industry because of its ease of implementation.
However, tuning PID gains in order to achieve the desired performance is a fairly challenging problem.
In general, experimental methods involving trial and error are used to tune these gains \cite{bo2016quadrotor, wang2016dynamics}.

There exist several methods to tune PID gains in quadcopters to achieve better performance in stability, transient response, and steady-state accuracy.
For example, the classic Ziegler-Nichols method\cite{ziegler1942optimum} was used in \cite{he2014simple}.
LQR control can also be implemented to obtain optimized PID gains by solving the Riccati equation \cite{mukhopadhyay1978pid}.
LQR-based tuning methods for quadcopters are discussed further in \cite{argentim2013pid,alkhoori2017pid}.
In \cite{bolandi2013attitude}, PID gains are determined using the direct synthesis method \cite{smith1997principles},
which is also an optimization-based method with constant variation in time rate.
Robust PID control for quadcopters is discussed in \cite{garcia2012robust}, which analyzes the sensitivity to achieve robustness from uncertainties like time delays incurred in actuation systems.
However, there is little or no work on algorithmically tuning PID gains to reject wind disturbances experienced in real-time flight.

In this work, we propose an $\mathcal{H}_2$ optimal PID controller that can reject the wind disturbance, and compare the performance of the proposed controller with the existing LQR based tuning method \cite{argentim2013pid}.


The rest of the paper is organized as follows. We first present the details of the quadcopter model in Section \ref{sec:quadModel} followed by a brief discussion on the conventional $\mathcal{H}_2$ optimal control framework in Section \ref{sec:h2control}. In Section \ref{sec:PID tune}, we discuss the proposed $\mathcal{H}_2$-optimal method for tuning the PID gains. Simulation results obtained using the proposed controller are presented and compared with the LQR-based controller in Section \ref{sec:res}. Concluding remarks and future research directions are provided in Section \ref{sec:concl}.

\section{Quadcopter models} \label{sec:quadModel}
In this section, we discuss quadcopter configuration and the mathematical model relevant to this work.
Detailed mathematical models for a quadcopter can be found in the references mentioned in Section \ref{sec:intro}.

For the purpose of this paper, we adopt the quadcopter model linearized about the hover state discussed in \cite{jivrinec2011stabilization}.
The lateral, longitudinal, directional, and vertical controllers can be decoupled in this model as shown in Fig. \ref{Fig:control_structure}.
The controller designed using this linearized model performs well in the nonlinear model. We compare the results of the proposed controller with the one based on LQR from \cite{argentim2013pid} which also uses the same dynamics model.

\subsection{Configuration}
\begin{figure}[ht]
\centering
\includegraphics[width=0.44\textwidth]{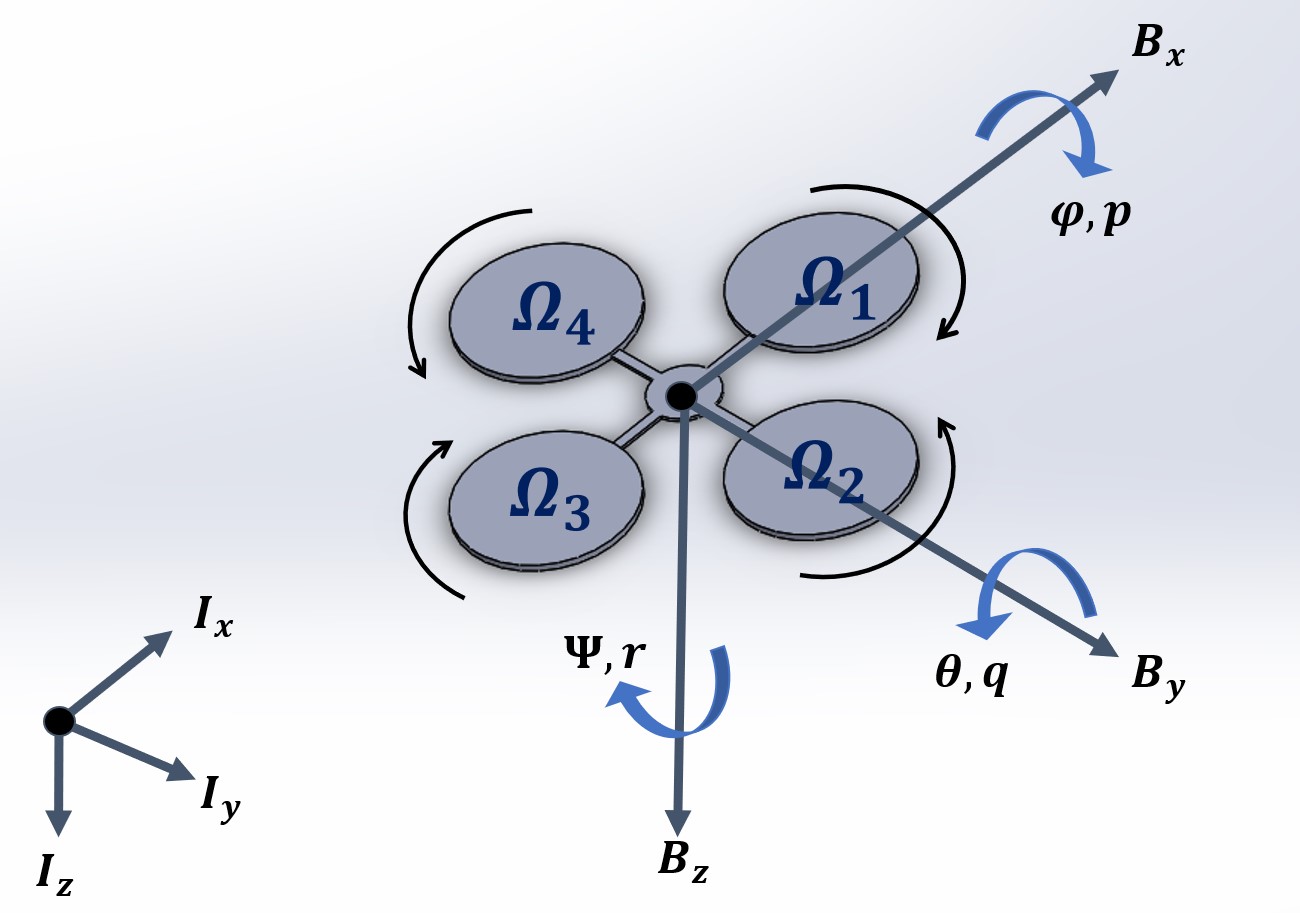}
\caption{The quadcopter configuration and frames of reference.}
\label{Fig:Frame}
\end{figure}

A quadcopter configuration is presented in Fig. \ref{Fig:Frame}, which has four motors and propellers that generate force and torque at each position.
Here, $\Omega$ is the rotor angular velocity used to control the vehicle.

\subsection{Dynamics} \label{sec:dyn_model}
Newton-Euler equations are used for representing the rigid body dynamics of the quadcopter.
The 6-DoF dynamics model is shown in Fig. \ref{Fig:Frame} with the \textit{Inertial frame} ($I_x, I_y, I_z$) and \textit{Body frame} ($B_x, B_y, B_z$).
$\phi, \theta, \psi$ are Euler angles in the inertial frame, and $p,q,r$ are angular velocities in the body frame about each axis.
These 6 variables are states for the \textit{rotational} motion.
Similarly, $x, y, z$ are the position coordinates in the inertial frame, and $u, v, w$ are velocities in the body frame about each axis. These 6 variables are states for \textit{translational} motion.

For the brevity of discussion, equations of motion for the quadcopter are omitted from this paper. However, we would like to note that the vehicle can be controlled with four inputs, $U_i$, which are combinations of four rotor angular velocities, $\Omega_i$, given by
\begin{subequations}
\label{eqn:input}
\begin{align}
    \text{Altitude control:}\quad U_1 &= b \ ({\Omega_1}^2 + {\Omega _2}^2 + {\Omega _3}^2 +{\Omega _4}^2)  \\
    \text{Roll control:}\quad U_2  &= b \ ({\Omega_2}^2 - {\Omega_4}^2) \\
    \text{Pitch control:}\quad U_3 &= b \ ({\Omega_1}^2 - {\Omega_3 }^2) \\
    \text{Yaw control:}\quad U_4 &= d \ ({\Omega_1}^2 - {\Omega_2 }^2+ {\Omega_3}^2-{\Omega_4}^2)
\end{align}
\end{subequations}
with thrust coefficient $b$, and drag coefficient $d$.

\begin{figure*}[t]
\centering
\includegraphics[scale=0.35]{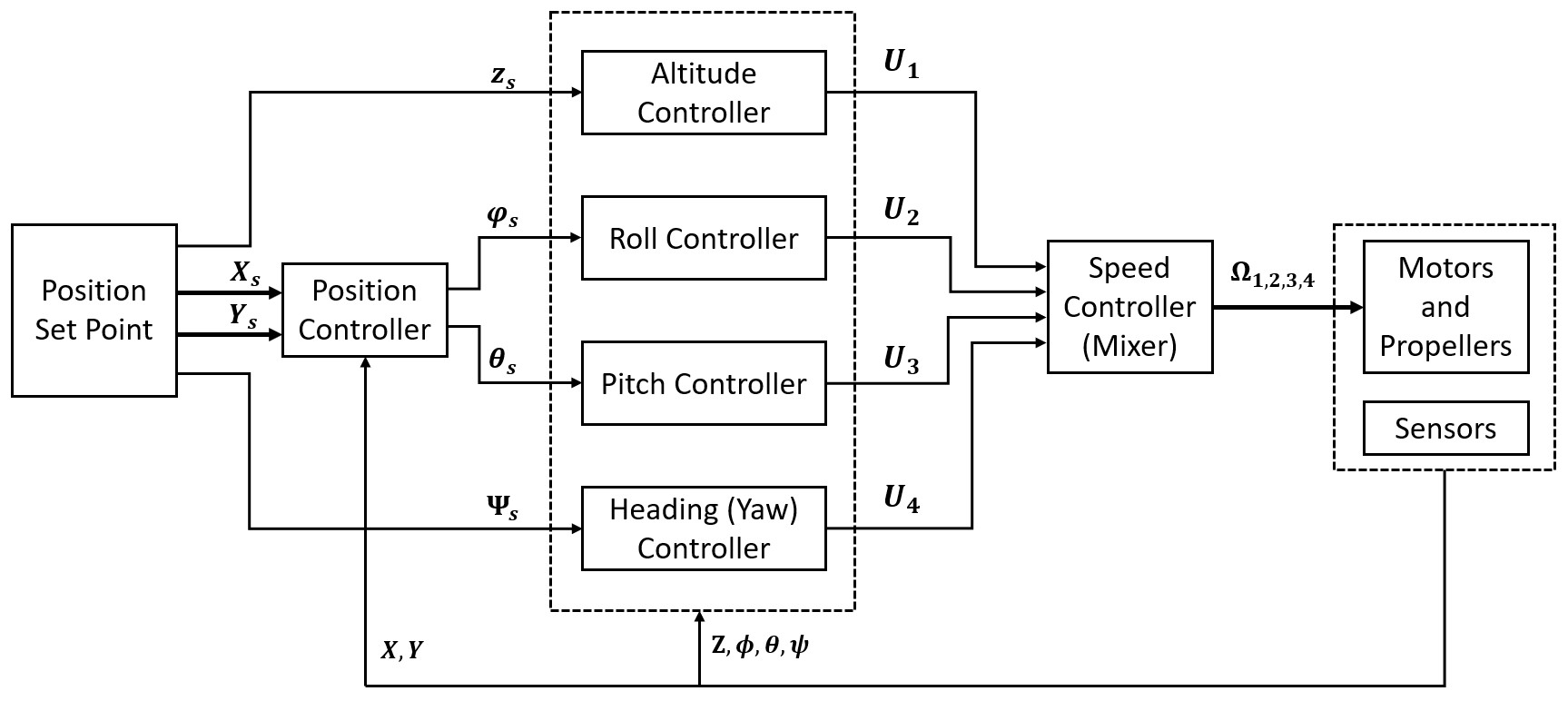}
\caption{The quadcopter control system: The complex nonlinear coupled model are decomposed into the four independent control subsystems with input combinations $U_i$, $i=1,2,3,4$.}
\label{Fig:control_structure}
\end{figure*}

Therefore, the complex nonlinear coupled model is decomposed into four control subsystems with input combinations (\ref{eqn:input}), as illustrated in Fig. \ref{Fig:control_structure}.
This allows us to consider each subsystem as a SISO (Single Input Single Output) system instead of a MIMO (Multi Input Multi Output) system to control the vehicle.
The control variables $U_i$ are calculated independently from each of the four control subsystems and fed into the mixer, which then calculates the individual rotor angular velocities $\Omega_i$.
We focus on the altitude control subsystem in this paper.


\subsection{Linearized model} \label{sec:linearized_sys}
We will use the linearized model to design the controller for the altitude control in hover state.
The following equations are considered for the vertical motion of the quadcopter:
\begin{subequations} \label{eqn:Vertical_Lin_sys}
\begin{align}
    \Dot{z} &= w_v \\
    \Dot{w}_v &= -2 \ \Omega_0 \ \frac{b}{m} \ (\Omega_1 +\Omega_3 -\Omega_2 -\Omega_4) + w\\
    \Dot{\Omega}_i &= -10 \ \Omega_i + 7u, \ i=1,2,3,4 \label{eqn:motor}
\end{align}
\end{subequations}
where $z$ is the altitude, $w_v$ is vertical speed, $w$ is disturbance, $b$ is the thrust coefficient (=$1.5108 \times 10^{-5} \ kgm $), $m$ is mass (=$1.07 \ kg$), $u$ is motor input as PPM (Pulse Position Modulation) signal. The numerical coefficients of $\Omega_i$ and $u$ in the above equations follow from the linearized transfer function of motor at hover state. The set of equations (\ref{eqn:Vertical_Lin_sys}) can represented in the state space form as
\begin{subequations} \label{eqn:Vertical_Lin_sys_ss}
\begin{align}
     \Dot{\vo{x}}(t) &= \vo{A} \vo{x}(t) + \vo{B}_{u} \vo{u}(t) + \vo{B}_w \vo{w}(t)\\
     \vo{y}(t) &= \vo{C} \vo{x}(t)
 \end{align}
\end{subequations}
with states as
\begin{align}
\vo{x}:= \begin{pmatrix} z, & w_v, & \Omega_{1},&\Omega_{2},&\Omega _{3}, &\Omega_{4} \end{pmatrix} ^T
\end{align}
and
\begin{align*}
    \vo{A}
    &=\begin{bmatrix} 0 & 1 & 0 & 0 & 0 & 0 \\
                        0 & 0 & -0.0106  & 0.0106  & -0.0106  & 0.0106  \\
                        0 & 0 & -10 & 0 & 0 & 0 \\
                        0 & 0 & 0 & -10 & 0 & 0 \\
                        0 & 0 & 0 & 0 & -10 & 0 \\
                        0 & 0 & 0 & 0 & 0 & -10
                \end{bmatrix},\\
    \vo{B}_u &=\begin{bmatrix} 0 & 0 & 7 & -7 & 7 & -7
                \end{bmatrix}^T,\\
    \vo{B}_w &=\begin{bmatrix}  0 & 1 & 0 & 0 & 0 & 0
                \end{bmatrix}^T,\\
    \vo{C} &= \begin{bmatrix}  1 & 0 & 0 & 0 & 0 & 0\\  0 & 1 & 0 & 0 & 0 & 0   \end{bmatrix}.
\end{align*}
We consider the system given by (\ref{eqn:Vertical_Lin_sys_ss}) to design the controller using LQR and $\mathcal{H}_2$ optimal control theory, which is discussed next.

\section{LQR and $\mathcal{H}_2$ Optimal Control} \label{sec:h2control}
In this section, we present very briefly, the necessary background for $\mathcal{H}_2$ optimal control theory for linear systems. Additionally, for comparison, LQR theory is also presented.
\subsection{Linear dynamic system}
We consider the following linear system,
\begin{subequations}\label{eqn:lin_system}
\begin{align}
\Dot{\vo{x}}(t) &=  \vo{A} \vo{x}(t) + \vo{B}_w \vo{w}(t) + \vo{B}_u \vo{u}(t)\\
\vo{z}(t) &= \vo{C}_z \vo{x}(t) + \vo{D}_u \vo{u}(t)\\
\vo{y}(t) &= \vo{C}_y\vo{x}(t)
\end{align}
\end{subequations}

where $\vo{x} \in \mathbb{R}^{n}$, $\vo{y} \in \mathbb{R}^{l}$, $\vo{z} \in \mathbb{R}^{m}$ are respectively the state vector, the measured output vector, and the output vector of interest respectively.
Variables $\vo{w} \in \mathbb{R}^{p}$ and $\vo{u} \in \mathbb{R}^{r}$ are the disturbance and the control vectors, respectively.

We are interested in designing a full state feedback controller for the system given by  (\ref{eqn:lin_system}), i.e.,
\begin{align} \label{eqn:control_law}
\vo{u}(t) = \vo{K} \vo{x}(t),
\end{align}
such that the closed loop system is stable and the effect of the disturbance is attenuated to the desired level.

\subsection{LQR optimal control}
The linear quadratic regulator (LQR) is a method used to determine the state feedback gain $K_{LQR}$. This controller is designed to minimize the cost function, $J$, defined as
\begin{align} \label{eqn:LQR_cost}
J = \int ^{\infty} _{0} (\vo{x}^T \vo{Q} \vo{x} + \vo{u}^T \vo{R} \vo{u}) dt
\end{align}
where $\vo{Q}\geq \vo{0}$ and $\vo{R} > \vo{0}$ are symmetric weighting matrices.
These matrices are main design parameters for defining the the control objective such that the state error and control energy is minimized.
The LQR problem can be converted to the LMI (Linear Matrix Inequality) form as given by the following theorem.

\textbf{Theorem 1 (LQR Optimal Control)} \cite{duan2013lmis} : The following two statements are equivalent:
\begin{enumerate}
    \item A solution $\vo{K}_{LQR}$ to the LQR controller exists.
    \item $\exists$ a matrix $\vo{Y}$, a symmetric matrix $\vo{W}$, and a symmetric matrix $\vo{Y}=\vo{P}^{-1}$ such that:
\end{enumerate}
\begin{align}
    \vo{AY}+ \vo{Y} \vo{A}^T +  \vo{W}^T \vo{B}_u ^T + \vo{B_{u}} \vo{W} &+ \vo{YQY} + \vo{W}^T \vo{RW} < 0  \label{LQR_LMI}
\end{align}
The optimal LQR control gain, $\vo{K}_{LQR}$, is determined by solving the following optimization problem.
\begin{align}
\min_{\vo{P},\vo{W},\vo{Y}} \quad  \textbf{trace}\  (\vo{P})  \quad \text{subject to (\ref{LQR_LMI}). \nonumber}
\end{align}
The gain $\vo{K}_{LQR}$ is recovered by $\vo{K}_{LQR} = \vo{W} \vo{Y}^{-1} $. This optimal gain minimizes the cost function (\ref{eqn:LQR_cost}).


\subsection{$\mathcal{H}_2$ Optimal Control}

With the linear system (\ref{eqn:lin_system}) and control law (\ref{eqn:control_law}), the $\mathcal{H}_2$ control closed-loop has the following form,
\begin{subequations} \label{control_loop}
\begin{align}
    \Dot{\vo{x}}(t) &= ( \vo{A} + \vo{B}_u \vo{K}) \vo{x}(t) + \vo{B}_z \vo{w}(t),\\
    \vo{z}(t) &= (\vo{C}_z + \vo{D}_u \vo{K} ) \vo{x}(t),
\end{align}
\end{subequations}
Therefore, the influence of the disturbance $\vo{w}$ on the output $\vo{z}$ is determined in frequency domain as
\begin{align}
\vo{z} = \vo{G}_{z w}(s) \vo{w}(s)
\end{align}
where $\vo{G}_{zw}(s)$ is the transfer function from the disturbance $\vo{w}$ to the output $\vo{z}$ given by
\begin{align} \label{errorTF}
\vo{G}_{z w}(s) = \vo{C}_z  (\vo{C}_z + \vo{D}_u \vo{K})[s \vo{I} -( \vo{A} + \vo{B}_u \vo{K})]^{-1} \vo{B}_w .
\end{align}



The problem of $\mathcal{H}_2$ optimal control design is then, given a system (\ref{errorTF}) and a positive scalar $\gamma$, find a matrix $\vo{K}_{\mathcal{H}_2}$ such that
\begin{align}\label{min1}
\|\vo{G}_{zw}(s)\|_{2} < \gamma.
\end{align}
The formulation to obtain $\vo{K}_{\mathcal{H}_2}$ is given by the following theorem.\\
\textbf{Theorem 2 ($\mathcal{H}_2$ Optimal Control)} \cite{duan2013lmis, apkarian2001continuous} : The following two statements are equivalent:
\begin{enumerate}
    \item A solution $\vo{K}_{\mathcal{H}_{2}}$ to the $\mathcal{H}_2$ controller exists.
    \item $\exists$ a matrix $\vo{W}$, a symmetric matrix $\vo{Z}$, and a symmetric matrix $\vo{X}$ such that:
\end{enumerate}
\begin{align}\label{H2_LMI}
\vo{AX}+\vo{B}_u \vo{W} +(\vo{AX}+\vo{B}_u \vo{W})^T + \vo{B}_w \vo{B}_w ^T &< 0  \nonumber\\
\begin{bmatrix}\nonumber
\vo{-Z}  &  \vo{C}_z \vo{X} + \vo{D}_z \vo{W} \\
\vo{*} & \vo{-X}
\end{bmatrix} &< 0\\
\textbf{trace}(\vo{Z}) &< \gamma^2
\end{align}
The minimal attenuation level $\gamma$ is determined by solving the following optimization problem
\begin{align}
\min_{\vo{W},\vo{X}, \vo{Z}} \gamma \quad \text{ subject to (\ref{H2_LMI}). }  \nonumber
\end{align}
The $\mathcal{H}_2$ optimal control gain is recovered by $\vo{K}_{\mathcal{H}_2}= \vo{W} \vo{X}^{-1}$. This optimal gain ensures that the closed-loop system is asymptotically stable and attenuates the disturbance. 





\section{$\mathcal{H}_2$ PID tuning method} \label{sec:PID tune}
In this section, we present the proposed PID tuning method based on $\mathcal{H}_2$ framework, which is an extension of the work in \cite{mukhopadhyay1978pid}.

The control input $\vo{u}$ from a PID controller is given by
\begin{align}
    \vo{u} = - K_P \ y - K_I \int^t_0 y \ dt - K_D \  \dot{y}
\end{align}
where $K_P, K_I$ and $K_D$ are proportional, integral, and derivative feedback gains respectively.
Eliminating $y$ using linear system equations (\ref{eqn:lin_system}) yields the extended form of the control law
\begin{align}
     \vo{u} &= - K_P \ \vo{Cx} - K_I \int^t_0 y \ dt \nonumber \\
     & \quad - K_D \ \vo{C} (\vo{Ax} + \vo{B_u \vo{u}} +\vo{B_w w})\nonumber\\ \nonumber
     &= - ( K_{P} \vo{C} + K_{D} \vo{CA}) x -  K_D \vo{C B_u \vo{u}} - K_D \vo{C B_w w} \nonumber \\
     &\ \hspace{60mm}- K_I \int^t_0 y \ dt \nonumber\\
     &= - (\vo{I}+ K_D \vo{C B})^{-1} ( K_{P}\vo{C} + K_{D} \vo{CA}) \ \vo{x} \nonumber\\
     & \ \hspace{20mm}  - (\vo{I}+ K_D \vo{C B})^{-1} K_D \vo{C B_w} \ \vo {w} \nonumber \\
     & \ \hspace{30mm} - (\vo{I}+ K_D \vo{C B})^{-1} K_I \int^t_0 y \ dt
\end{align}
We can rewrite this equation as
\begin{align} \label{eqn:PID_control_law_aug}
    \vo{u} = - \vo{Mx} - \vo{Nw} - \vo{L}  \int^t_0 \vo{y} \ dt
\end{align}
where
\begin{subequations}
\begin{align}
    \vo{M} &= (\vo{I}+ K_D \vo{C B})^{-1} ( K_{P}\vo{C} + K_{D} \vo{CA}) \label{eqn:M} \\
    \vo{N} &= (\vo{I}+ K_D \vo{C B})^{-1} K_D \vo{C B_w} \label{eqn:N}\\
    \vo{L} &= (\vo{I}+ K_D \vo{C B})^{-1} K_I\label{eqn:L}.
\end{align}
\end{subequations}
Note that the PID control law depends on signals from states ($x$), disturbance ($w$), and integration of the measurements ($\int^t_0 y \ dt$). Also, contribution of the disturbance signal $w$ to the control input $\vo{u}$ is affected by the gain $K_D$ in PID control.

Now, we can compare the $\mathcal{H}_2$ control law $\vo{u} = \vo{K}_{\mathcal{H}_2} \vo{x}$ with the PID control law (\ref{eqn:PID_control_law_aug}) to get the PID gains.
However, there are two more terms in the control law which are dependent on $w$ and $ \int^t_0 y \ dt$.
We can disregard the term associated with $w$ for the purpose of comparison, because $w$ is already attenuated in ${\mathcal{H}_2}$ control framework.
To handle the term associated with $ \int^t_0 y \ dt$, we define a new state, $\vo{\zeta}$, as
\begin{subequations}
\begin{align}
    \vo{\zeta} &:= \int^t_0 y \ dt \\   \dot{\vo{\zeta}} &= \vo{y} = \vo{Cx} .
\end{align}
\end{subequations}

We define the augmented state vector as $\vo{\Bar{x}}:=[ \vo{x} \ \vo{\zeta}]^T$, and the augmented system is represented in the state space form as
\begin{align}
\label{eqn:Aug_system}
\Dot{\Bar{\vo{x}}}(t) = \vo{\bar{A}} \vo{x}(t) + \vo{\Bar{B}}_{w} \vo{w}(t) + \vo{\Bar{B}}_{u} \vo{u}(t)
\end{align}
i.e.,
\begin{align}
    \begin{bmatrix}
        \dot{\vo{x}} \\ \dot{\vo{\zeta}}
    \end{bmatrix}
    =
    \begin{bmatrix}
        \vo{A} & \vo{0} \\ \vo{C} & \vo{0}
    \end{bmatrix}
    \begin{bmatrix}
        \vo{x} \\ \vo{\zeta}
    \end{bmatrix}
    +
     \begin{bmatrix}
        \vo{B}_u \\  \vo{0}
    \end{bmatrix}
    \vo{u}
    +
     \begin{bmatrix}
        \vo{B}_w \\  \vo{0}
    \end{bmatrix}
    \vo{w} \nonumber
\end{align}
Now, we can derive an optimal control law with $\mathcal{H}_2$ control theory for the augmented system as
\begin{align} \label{eqn:H2_control_law_aug1}
    \Bar{\vo{u}} &= - \Bar{\vo{K}}_{\mathcal{H}_2} \vo{\Bar{x}} = - [ \vo{K}_1 \ \vo{K}_2 ]  \begin{bmatrix} \vo{x} \\ \vo{\zeta}\end{bmatrix}
\end{align}
Let us rewrite the PID control law ({\ref{eqn:PID_control_law_aug}}) for the comparison  as
\begin{align} \label{eqn:PID_control_law_aug1}
    \vo{u} = - \vo{Mx} - \vo{L \zeta} = - [ \vo{M} \ \vo{L} ]  \begin{bmatrix} \vo{x} \\ \vo{\zeta}\end{bmatrix}.
\end{align}
Now, we can directly compare the two equations (\ref{eqn:H2_control_law_aug1}), (\ref{eqn:PID_control_law_aug1}) to get
\begin{align} \label{eqn:compare_control_law1}
    \vo{M} =  \vo{K}_1 \ \text{and}  \  \vo{L} = \vo{K}_2.
\end{align}

Once we know $\vo{M}$ and $\vo{L}$, equations (\ref{eqn:M}) and (\ref{eqn:L}) can be solved for $K_P$, $K_D$, and $K_I$ as
\begin{subequations}
\begin{align}\label{eqn:compare_control_law2}
     [K_P \ K_D] &= \vo{M} \begin{bmatrix} \vo{C}\\ \vo{CA} - \vo{CB M}
     \end{bmatrix}^{-1}\\
     K_I &= (\vo{I}+ K_D \vo{C B}) \vo{L}
\end{align} \label{eqn:h2_pid_gains}
\end{subequations}
The PID gains obtained by (\ref{eqn:h2_pid_gains}) result in the $\mathcal{H}_2$ optimal PID controller.
\section{Results} \label{sec:res}

\subsection{Simulation set up}
\begin{figure}[ht]
\centering
\includegraphics[width=0.4\textwidth]{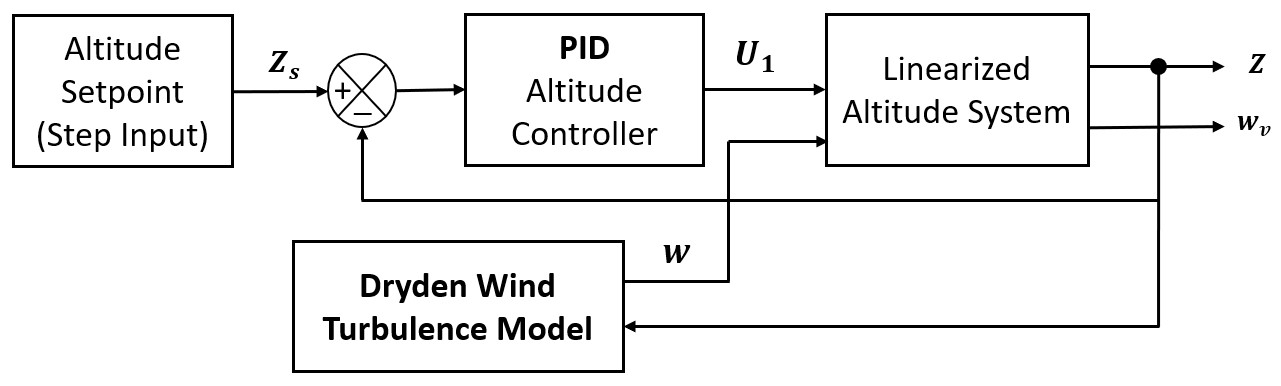}
\caption{Simulation structure for vertical altitude control}
\label{Fig:simulation}
\end{figure}

\begin{figure}[ht]
\centering
    \includegraphics[width=0.4\textwidth]{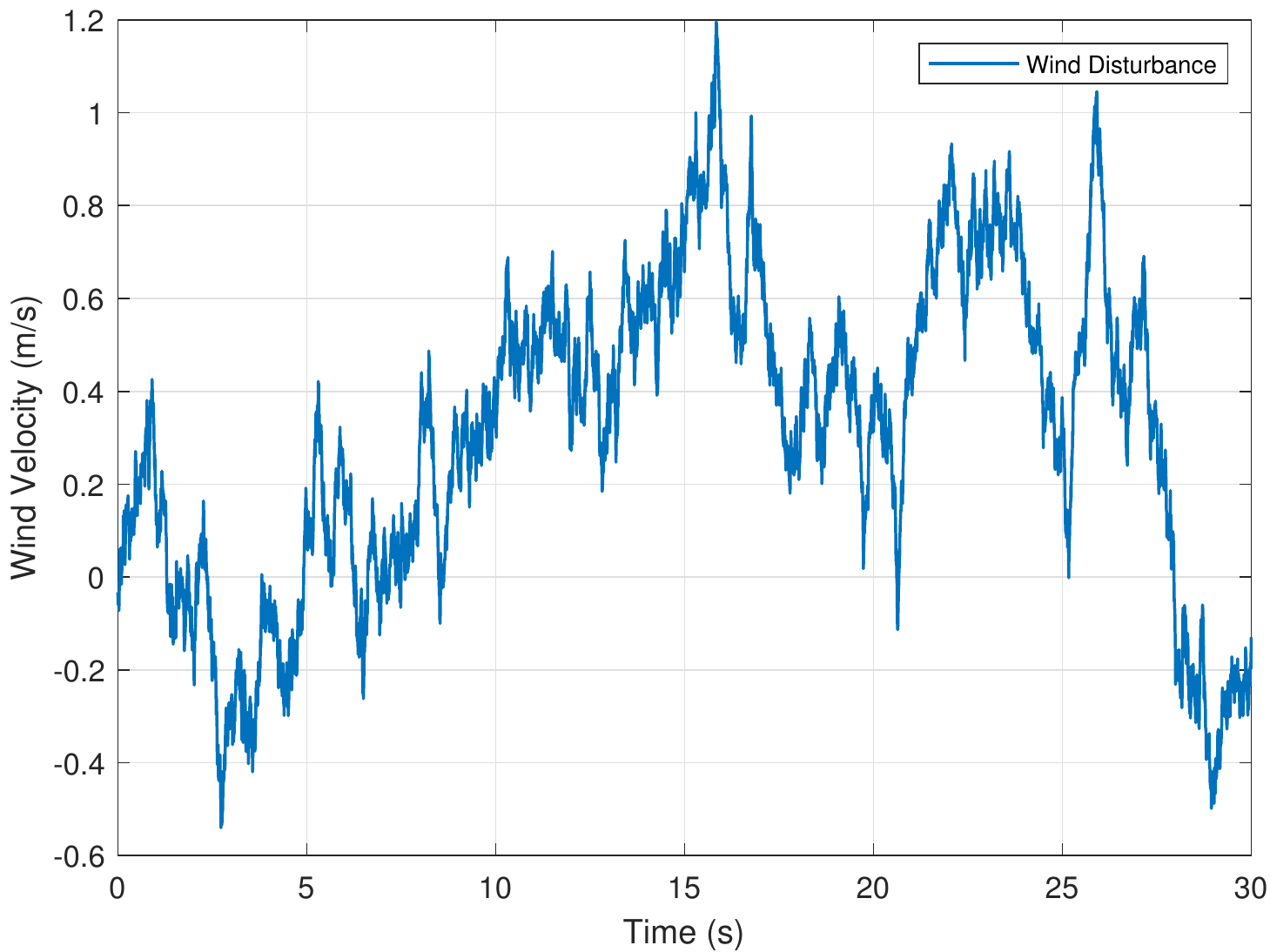}
    \caption{Wind disturbance along the Z axis generated by the Dryden wind turbulence model in the Simulink software.}
\label{Fig:wind}
\end{figure}

The proposed  $\mathcal{H}_2$ optimal PID controller is applied to the vertical altitude system (\ref{eqn:Vertical_Lin_sys}), and its performance is compared with the LQR based PID controller. The comparison is done in terms of control input, time response, and the amount of wind disturbance rejection, in a MATLAB based simulation environment, as shown in Fig. \ref{Fig:simulation}. The Dryden wind turbulence model was used to generate the wind disturbance in the Simulink software. The generated wind disturbance is 5 $m/s$ from north and component of $z$ direction shown in Fig. \ref{Fig:wind}.

\subsection{Simulation results} \label{sec:TF_SISO}

As discussed below, we consider two cases to analyze the performance of the proposed $\mathcal{H}_2$ optimal PID control algorithm for the vertical altitude system presented in \S \ref{sec:linearized_sys}.

\textbf{Case I: Vertical Velocity Control}\label{vel_process} -- In this case, we consider the vertical velocity control problem with the wind disturbance.
To solve the control problem, we've done minimal realization of the linearized model (\ref{eqn:Vertical_Lin_sys}) with the input as PPM signal and the output as vertical velocity.
The transfer function from input to output for this case is given by
\begin{align} \label{eqn:TF_vel}
    \vo{G}_{velocity} = \frac{-0.2968}{s(s+10)}.
\end{align}

This transfer function is represented in the state space form with disturbance as
\begin{subequations} \label{eqn:Vel_sys}
\begin{align}
    \begin{bmatrix}
        \dot{x}_1 \\ \dot{x}_2
    \end{bmatrix}
    &=
    \begin{bmatrix}
        -10 & 0 \\ 1 & 0
    \end{bmatrix}
    \begin{bmatrix}
        x_1 \\ x_2
    \end{bmatrix}
    +
     \begin{bmatrix}
        1 \\  0
    \end{bmatrix}
    \vo{u}
    +
     \begin{bmatrix}
        0 \\  1/0.2968
    \end{bmatrix}
    \vo{w} \\
y &=\begin{bmatrix} 0 & -0.2968
    \end{bmatrix} x   .
\end{align}
\end{subequations}
Here, scaled disturbance matrix is multiplied with $w$, since state $x_2$ is the scaled velocity in the minimal realization of the system.

The augmented system of (\ref{eqn:Vel_sys}) follows from (\ref{eqn:Aug_system}) as
\begin{align} \label{eqn:Vel_aug_sys}
    \begin{bmatrix}
        \dot{x}_1 \\ \dot{x}_2 \\ \dot{\zeta}
    \end{bmatrix}
    &=
    \begin{bmatrix}
        -10 & 0 & 0\\ 1 & 0 & 0 \\ 0 & \frac{-1}{0.2968} & 0
    \end{bmatrix}
    \begin{bmatrix}
        x_1 \\ x_2 \\ \zeta
    \end{bmatrix}
    +
     \begin{bmatrix}
        1 \\  0 \\ 0
    \end{bmatrix}
    \vo{u}
    +
     \begin{bmatrix}
        0 \\  \frac{1}{0.2968} \\ 0
    \end{bmatrix}
    \vo{w}
\end{align}
And, we set the performance output vector as
\begin{align} \label{eqn:performance_vel}
    \vo{z} &=\begin{bmatrix} \ 0 & 0 & 100 \ \end{bmatrix}\vo{x} + \vo{u}
\end{align}

We can determine the $\mathcal{H}_2$ optimal solution for the augmented system (\ref{eqn:Vel_aug_sys}) from (\ref{H2_LMI}) with input weight $W_u$ = $0.01$. Herein, $W_u$ is used to reflect the restrictions on the actuator signals.

And, we obtain the optimized PID gains with (\ref{eqn:compare_control_law1}), (\ref{eqn:h2_pid_gains}) as
\begin{align}
\text{$\mathcal{H}_2$:} \ K_P=-1170.8, \ K_I=-1, \ K_D=-115.1. \ \end{align}
Note that, as expected, the large magnitude of $K_D$ gain is obtained to counter the wind disturbance.
For the LQR tuned PID, we use $\vo{Q} = diag \ ([0 \ 10000 \ 10000])$ and $\vo{R}=1$, and it results in the following PID gains
\begin{align}
\text{LQR:} \ K_P=-354.1, \  K_I=-1, \ K_D=-25.6.
\end{align}

\begin{figure}[!ht]
\centering
    \includegraphics[width=0.44\textwidth]{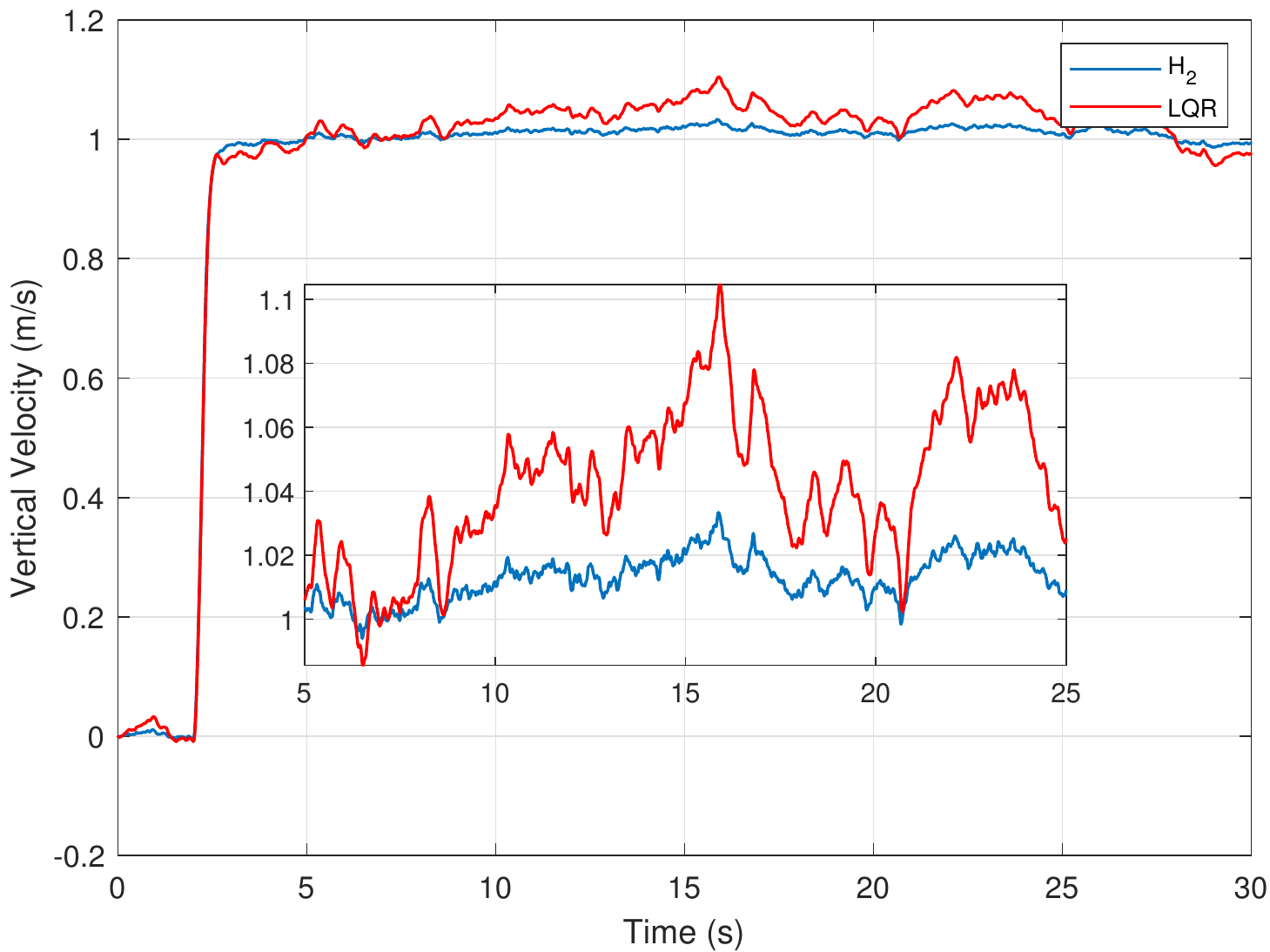}
    \caption{Step response of vertical velocity control.}
\label{Fig:vel_step}
\end{figure}
\begin{figure}[!ht]
\centering
    \includegraphics[width=0.44\textwidth]{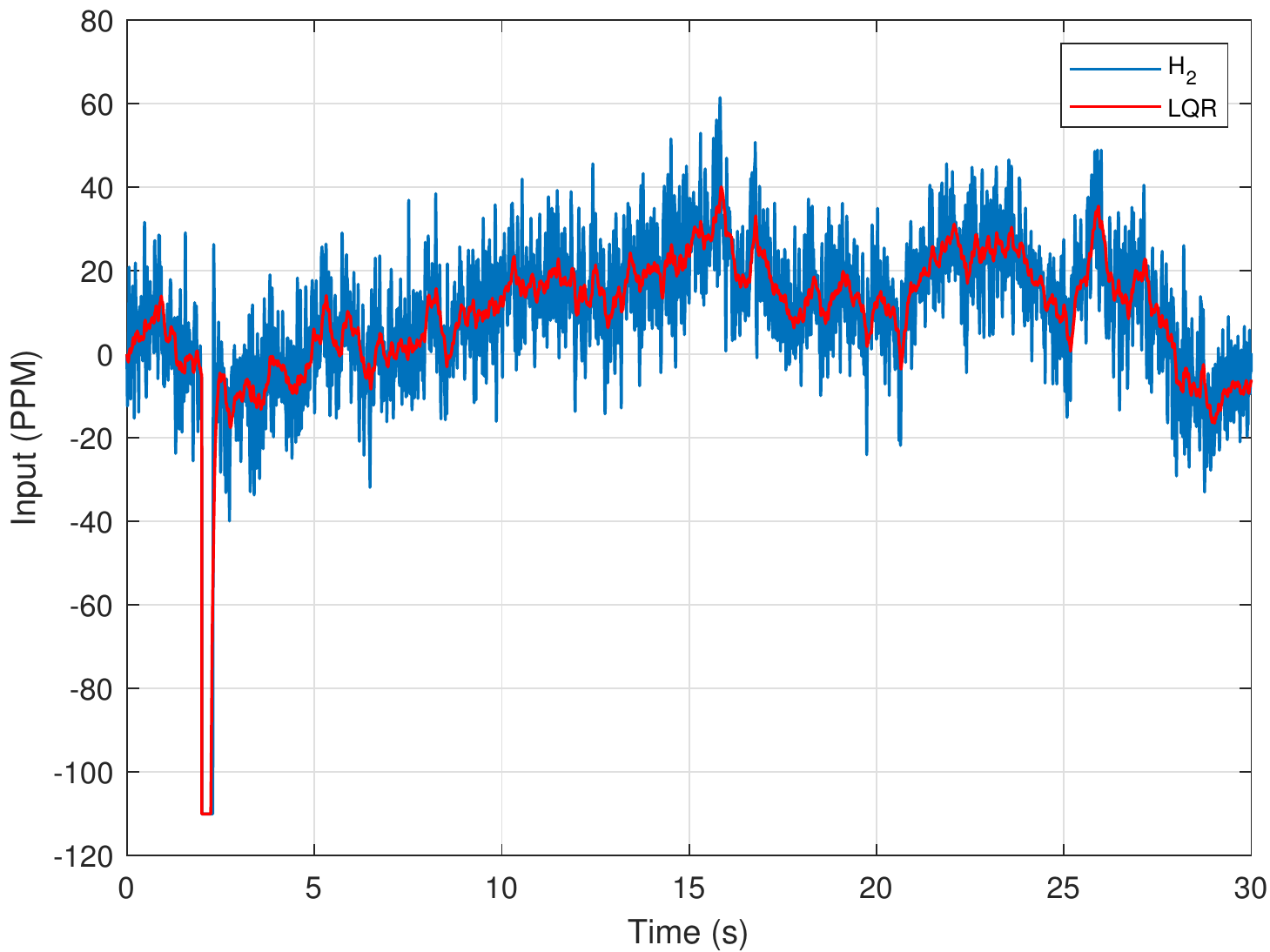}
    \caption{Control input for vertical velocity control.}
\label{Fig:vel_input}
\end{figure}

The simulation results are shown in Fig. \ref{Fig:vel_step}. and Fig. \ref{Fig:vel_input}. As shown in Fig. \ref{Fig:vel_step}, we observe that the $\mathcal{H}_2$ tuned PID controller demonstrates better wind disturbance rejection than the LQR tuned PID controller with similar response time.

When we consider energy consumption, as shown in Fig. \ref{Fig:vel_input}., we observe that the $\mathcal{H}_2$ tuned PID controller requires higher variance in control input to count wind disturbance.
However, $\mathcal{H}_2$ method use slightly higher input energy than the LQR tuned PID controller when we check its mean in Table \ref{table:Comp_vel_Input}.

\begin{table}[!ht]
\caption{Comparison of vertical velocity control Input } \label{table:Comp_vel_Input}
\begin{center}
\renewcommand{\arraystretch}{1.5}
\begin{tabular}{|c||c|c|}
\hline
Tuning algorithm & Mean (PPM) & Covariance (PPM)\\
\hline \hline
$\mathcal{H}_{2}$-PID & 9.3751 & 294.5 \\
\hline
LQR-PID & 9.3909  & 267.9\\
\hline
\end{tabular}
\end{center}
\end{table}

\textbf{Case II: Vertical Position Control} --
In this case, we consider the vertical position control problem with the wind disturbance. Similar to the previous case, we get the transfer function:
\begin{align} \label{eqn:TF_pos}
    \vo{G}_{position} = \frac{-0.2968}{s^{2}(s+10)}.
\end{align}
We set the input weight $W_u$ = $0.1$ and performance output $\vo{z}$ is defined as:
\begin{align}
    \vo{z} &=\begin{bmatrix}
        0 & 0 & 100 & 1000
    \end{bmatrix} \vo{x} + \vo{u} .
\end{align}

\begin{figure}[t]
\centering
    \includegraphics[width=0.44\textwidth]{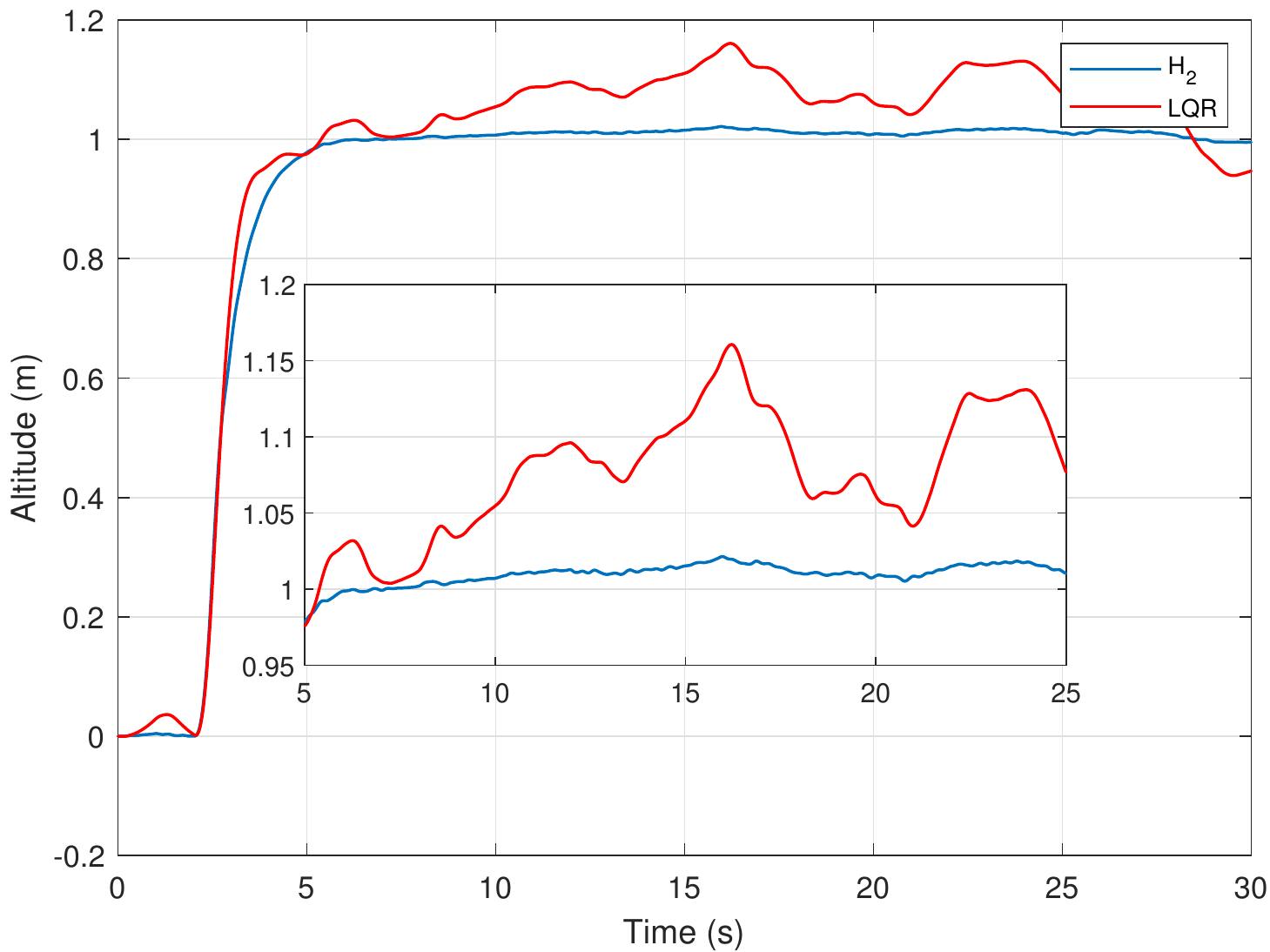}
    \caption{Step response of vertical position control}
\label{Fig:position_step}
\end{figure}

\begin{figure}[t]
\centering
    \includegraphics[width=0.44\textwidth]{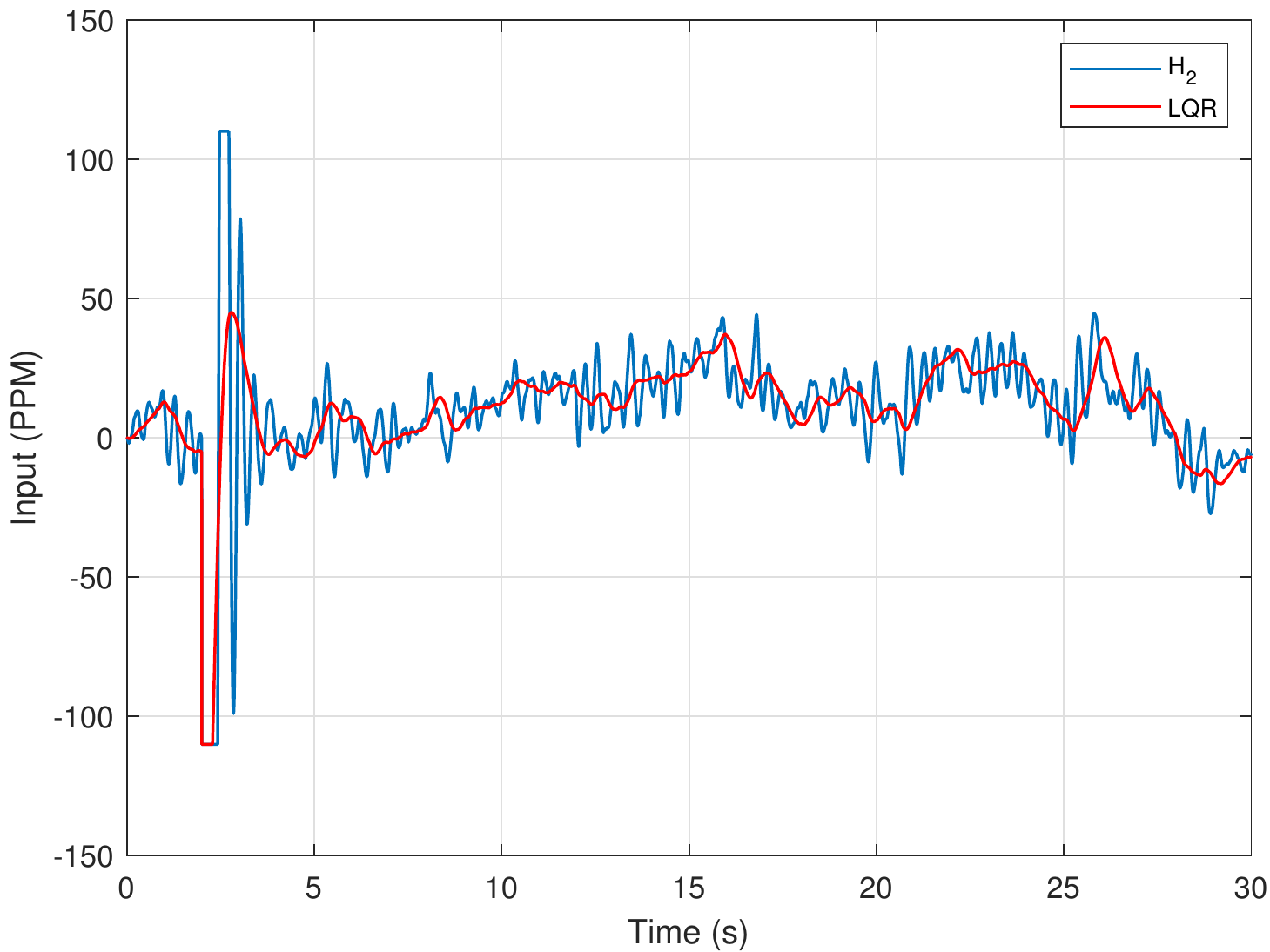}
    \caption{Control input for vertical position control}
\label{Fig:position_input}
\end{figure}

The PID gains obtained using $\mathcal{H}_2$ optimal tuning are:
\begin{align}
\text{$\mathcal{H}_2$:} \ K_P=-1370, \ K_I=-1, \ K_D=-881.7.
\end{align}
For the LQR case, we use $\vo{Q}=diag  ([0 \ 0 \ 1000 \ 10000])$ and $\vo{R}=1$, and we obtain the following PID gains :
\begin{align}
\text{LQR:} \ K_P=-192.6, \  K_I=-1, \ K_D=-128.7.
\end{align}
The simulation results are shown in Fig. \ref{Fig:position_step}. and Fig. \ref{Fig:position_input}. Similar to the previous case, we observe that $\mathcal{H}_2$ tuned PID controller rejects wind disturbance better than the LQR tuned PID controller, with similar response time, as shown in Fig. \ref{Fig:position_step}. Again, $\mathcal{H}_2$ tuned PID controller requires slightly higher control energy than the LQR tuned PID controller as shown in Table \ref{table:comp_position_Input}.

\begin{table}[h]
\caption{Comparison of vertical position control input } \label{table:comp_position_Input}
\begin{center}
\renewcommand{\arraystretch}{1.5}
\begin{tabular}{|c||c|c|}
\hline
Tuning algorithm & Mean (PPM) & Covariance (PPM)\\
\hline \hline
$\mathcal{H}_{2}$-PID & 10.4889 & 513.3045 \\
\hline
LQR-PID & 10.4681  & 317.9473\\
\hline
\end{tabular}
\end{center}
\end{table}


\section{Conclusions} \label{sec:concl}

This paper presented a new optimized PID control algorithm for quadcopter systems to counter wind disturbance, based on $\mathcal{H}_2$ optimal control theory. We showed that the proposed $\mathcal{H}_2$ optimal PID controller rejects the wind disturbance better than the existing LQR tuned PID controller. Since all UAVs are affected by wind disturbance in the real world flight environments, the ability of the proposed tuning method to reject these disturbances makes it very attractive for designing PID controllers.
This work considered models in the continuous time domain and results were obtained solely through simulation. Our future work will address discrete time systems and validation of the proposed controllers with experimental results.



\bibliographystyle{elsarticle-num}
\bibliography{root}

\end{document}